\documentclass[journal]{IEEEtran}
\usepackage{amsmath,amsfonts, amssymb}
\usepackage{algorithmic}
\usepackage{algorithm}
\usepackage{array}
\usepackage[caption=false,font=normalsize,labelfont=sf,textfont=sf]{subfig}
\usepackage{textcomp}
\usepackage{stfloats}
\usepackage{url}
\usepackage{verbatim}
\usepackage{graphicx}
\usepackage{cite}

\usepackage{multirow}
\usepackage{wrapfig}
\usepackage{subcaption} 
\usepackage{booktabs}
\usepackage{tikz}
\usetikzlibrary{patterns,fit,shapes.geometric,intersections,backgrounds,positioning,arrows.meta, calc, arrows,decorations.markings, bending}
\usepackage{pgfplots}
\pgfplotsset{compat=1.18}
\usepackage[nolist,nohyperlinks]{acronym}
\acrodef{GLM}[GLM]{graph neural network for line manifolds}
\acrodef{CT}[CT]{\emph{computed tomography}}
\acrodef{FBP}[FBP]{\emph{filtered backprojection}}
\acrodef{GNN}[GNN]{\emph{graph neural networks}}
\acrodef{CNN}[CNN]{\emph{convolutional neural networks}}
\acrodef{NN}[NN]{\emph{Neural Networks}}
\acrodef{PSNR}[PSNR]{\emph{peak signal-to-noise ratio}}
\acrodef{SSIM}[SSIM]{\emph{structural similarity}}

\newcommand{\Real}{\mathbb{R}}
\newcommand{\manifold}{\mathbb{M}}

\begin{document}

\title{Improving the Generalisation of \\ Learned Reconstruction Frameworks}

\author{Emilien Valat \and Ozan Öktem
\thanks{E.V and O.O are at KTH, Royal Institute of Technology, Sweden}
}

\markboth{Improving the Generalisation of Learned Reconstruction Frameworks}%
{Shell \MakeLowercase{\textit{et al.}}: A Sample Article Using IEEEtran.cls for IEEE Journals}


\maketitle

\begin{abstract}
Ensuring proper generalization is a critical challenge in applying data-driven methods for solving inverse problems in imaging, as neural networks reconstructing an image must perform well across varied datasets and acquisition geometries. In X-ray Computed Tomography (CT), convolutional neural networks (CNNs) are widely used to filter the projection data but are ill-suited for this task as they apply grid-based convolutions to the sinogram, which inherently lies on a line manifold, not a regular grid. The CNNs, unaware of the geometry, are implicitly tied to it and require an excessive amount of parameters as they must infer the relations between measurements from the data rather than from prior information.

The contribution of this paper is twofold. First, we introduce a graph data structure to represent CT acquisition geometries and tomographic data, providing a detailed explanation of the graph's structure for circular, cone-beam geometries. Second, we propose \ac{GLM}, a hybrid neural network architecture that leverages both graph and grid convolutions to process tomographic data.

We demonstrate that \ac{GLM} outperforms CNNs when performance is quantified in terms of structural similarity and peak signal-to-noise ratio, this desipte the fact that \ac{GLM} uses only a fraction of the trainable parameters. Compared to CNNs, \ac{GLM} also requires significantly less training time and memory, and its memory requirements scale better. Crucially, \ac{GLM} demonstrates robust generalization to unseen variations in the acquisition geometry, like when training only on fully sampled CT data and then testing on sparse-view CT data.
\end{abstract}


\section{Introduction}
\IEEEPARstart{T}{omography} is a non-destructive imaging technology where one measures the effect of probing an object with penetrating rays/waves from various directions.
Computational methods are then used to reconstruct the interior structure of object (\emph{image}) from such data.
The most common example of tomography is medical \ac{CT} where X-rays are used to image the interior anatomy of a subject.
The image is then the objects linear attenuation coefficient.

\subsection{Reconstruction problem}
We next formalize the computational task of tomographic image reconstruction. In doing so, it is common to mathematically represent the \emph{image} one seeks to recover with a real valued function $x \colon \Omega \to \Real$ with $\Omega \subset \Real^d$ ($d=2, 3$). 
Next, the \emph{acquisition geometry} in tomography is a precise specification of how data is collected. Mathematically, this corresponds to a scheme for sampling elements from some submanifold $\manifold$ of lines in $\Real^d$.

\emph{The reconstruction problem} is then the computational task of recovering the image $x$, which is a real-valued function on $\Omega \subset\Real^d$, from the \emph{sinogram}, which is a function $y \colon \manifold \to \Real$ that satisfies 
\begin{equation}\label{eq:InvProb}
    y = \mathcal{A}(x) + e.
\end{equation}
Hence, the sinogram is a function on lines that represents continuum data and $e$ is the corresponding (unknown) continuum observation error.
Finally, $\mathcal{A}$ is the \emph{forward operator} that models the data generated by an image in absence of noise and observation errors.
It is in most cases given as the solution to a radiative transport equation.
However, if raw sensor data has been recorded using appropriate instrumentation (anti-scatter grids, bow-tie filter,\ldots), then one can pre-process data so that it
can be interpreted as the ray transform on $\manifold$ of the object's linear attenuation coefficient, i.e., one can assume that
\[ 
\mathcal{A}(x)(\ell) = \int_\ell x(s) ds \quad\text{for $\ell \in \manifold$.}
\]

\subsection{Reconstruction methods}
Traditional approaches to image reconstruction entail a variety of frameworks. Analytic procedures such as \ac{FBP} rely on a closed-form solution of the inversion problem in a low-noise scenario. Iterative and variational approaches depend on solving the inverse problem using a gradient descent scheme, regularised by early stopping for the former or by a carefully chosen functional for the latter. Data-driven techniques build on these methods and enhance the robustness to the noise inherently occurring in natural processes and impossible to model accurately, and variability in the data by using \ac{NN}. Within this context, \ac{CNN} have proven to be efficient tools to enhance the image reconstruction procedure. 

The use of \ac{NN} for solving the reconstruction problem can be divided in two categories: the post-processing and learned reconstruction approaches. The former inputs an image already reconstructed and enhances it using dedicated architectures and loss functions. The latter inputs the sinogram, knowledge about the forward operator and its adjoint $\mathcal{A}^T$, and learns to perform a sinogram-to-image mapping. A historical example of learned post-processing is FBPConvNet \cite{jin_deep_2017}, and the Learned Primal-Dual \cite{adler_learned_2018} is a reference of the learned iterative frameworks. Post-Processing methods were a low-hanging fruit as the research on image processing (denoising, deblurring, inpainting, segmentation) is dense and benefits from inputs from several other fields. For instance,  \cite{kim_systematic_2024} compares $75$ post-processing methods to denoise low-dose CT (LDCT) images on the MAYO dataset \cite{mccollough_low_2020}, but note a problem with the generalisation capabilities of the test networks. Also, for LDCT, \cite{zhang_review_2024} compares the performance of $9$ post-processing networks among a reported number of $55$ methods. This wealth of papers focuses on the image denoising methods rather than image reconstruction, although the latter have been shown to outperform the former ones on real and simulated data \cite{kiss_benchmarking_2024, leuschner_quantitative_2021, moriakov_end--end_2023, genzel_near-exact_2022}. Although under the umbrella of image improvement, post-processing methods input an image, not a sinogram, which explains why they underperform compared to image reconstruction methods. 

The review \cite{xia_physics-model-based_2023} denotes 19 learned reconstruction algorithms and compares the performance of eight of them. According to their experience, learned iterative methods such as the LPD and FISTA-Net \cite{xiang_fista-net_2021} outperform all the other tested methods, whether model or denoising based. However, their conclusion highlights the need to improve the generalisability of the learned approaches. The same is drawn by \cite{li_comprehensive_2022}, where the authors note that the main problem in the 22 methods they study is their poor generalisation to new acquisition geometries, and recommend to add more prior knowledge in the networks and develop models that generalise to even small perturbations in the task at hand (``generalisation performance is poor in other tasks, even if the task is very similar to the original task.''). This is consistent with the conclusion of \cite{zhang_review_2024}. We are confident to say that the generalisation of learned reconstruction frameworks is a pressing issue in CT imaging. In this paper, we tackle it by focusing on sinogram data processing which is used in the learned reconstruction methods. 

Traditionally, CNNs have been used to filter sinogram data, as done in the LPD and FISTA-Net. The major issue with the use of CNNs to process sinograms is the dimension of their kernels. Indeed, if one uses kernels of the detector's dimension, there is no aggregation mechanism to leverage relations between neighbouring measurements. Alternatively, if one uses kernels of the sinogram's dimension, the kernels will expect a constant spacing for each of their dimension. Also, these kernels are not aware of the geometry, as there is no mechanism to encode the relation between measurements. Instead, they rely on an implicit parametrisation as they are trained against only one geometry. Recent developments in machine-learning architectures have shed light on the Transformers \cite{vaswani_attention_2017} and their spatial counterpart \cite{dosovitskiy_image_2021}, a family of neural-networks excelling at modelling long-range connexion within the data. As sinogram data is non-local, i.e pixels values depend on each other across the angular and detector dimensions in a geometry-dependent way, it is natural to want to use these architectures for realising sinogram-to-sinogram mappings in learned reconstruction frameworks, as underlined in \cite{xia_physics-model-based_2023, zhang_review_2024}.

There are three approaches reported to have used transformers in the projection domain. First, \cite{wang_dudotrans_2022} which addresses sparse-view CT. Their motivation is that sinograms are non-local and that the transformer should be able to pick-up long-range dependencies within them. The main problem with their approach is that there is no insurance that it scales for a larger number of measurements than the $144$ reported, as transformers are known to require a large number of parameters. Their explanation is also rather limited when it comes to the positional embedding of measurements, making their architecture tied to the number of measurements. Finally, they do not explain if they had to retrain a transformer for each acquisition geometry from scratch nor how to encode missing measurements which would negate the presumed advantage of Transformers for generalisation.

Second, \cite{yang_low-dose_2022}, which relies on a sinogram inner-structure loss. We see three issues in their approach. First, the inner-structure loss which assumes an availability of measurements exactly $\pi$ radians away from each other and a delicate selection of values along diagonal lines in the sinogram. The assumption of the availability of these measurements is alone strong for realistic experimental settings, but extracting information along diagonal lines is also difficult, as it involves an interpolation of values on a grid which has different resolutions in its $x$ and $y$ dimensions. One must also note that they do not give a recipe to select the said lines for a different geometry than the one they use. Finally, their transformer architecture uses both positional encoding and linear layers. The problem with the former is that positional encoding requires a notion of number of tokens and fixed acquisition geometry, hence impossibility to accommodate for the slightest variations. The problem with the latter is that linear layers are not adapted (this is not a bold claim) for image data, as the decade long research in CNN as shown. The absence of code to answer these problems hampers the value of their claims and makes us think that they used a parallel-beam geometry to compute their inner-structure loss and tailored their architecture to this fixed acquisition geometry. 

The final one is \cite{li_ddptransformer_2022}, which we tried to implement. To our surprise, their implementation assumed that the number of measurements and number of detector pixels were the same. It is an unreasonable assumption to imagine that any sinogram can be represented on a square grid. Also, when we computed the number of trainable parameters in their model for their settings from their open-source code, we found that they require $187.297$ parameters  for $512\times512$ sinograms, which indicates an impossibility to reasonably scale to real data. 

Although Transformers are hinted in most review papers as a promising architecture for image improvement in CT, our own unsuccessful experiences with Transformers for sinogram processing highlighted two significant issues with such approaches. First, tokenization. Consider Natural Language Processing, a field for which Transformers were developed and excel in. Transformers rely on mapping chunks of their input data into a vector of different dimension. Although the tokenization is learned, a sequence of words will be tokenized in a consistent manner as there is a finite number of possible tokens. However, in CT, if one is to consider a sinogram a sequence of ``measurement'' tokens, the amount of existing tokens is infinite, and predicting a 8 bytes integer pixel one at a time as done in \cite{parmar_image_2018} seems unreasonable for the sheer size of sinogram data. Second, positional encoding. In sentences, positional encoding of tokens is straightforward, but in CT, if a sinogram is understood as a sequence of measurements token, one issue arises immediately: the position of a token in a sequence must be parametrised by a minimal angular distance between measurements. Consider two acquisition geometries with the same trajectory, but different angular resolutions $R_1$ and $R_2$, with $R_1$ \textless $R_2$. The natural positional encoding puts two consecutive measurements one ``token'' distance away from each other. As such, encoding the angular positions in $R_1$ will be too crude for $R_2$, and encoding the positions based on $R_2$ will leave wide gaps (filled with $0$-arrays to represent missing tokens?) when working with the same Transformer on data with $R_1$ resolution. And the problem goes on for any other angular resolution not known a-priori. Note that this shortcoming could potentially be addressed with a continuous positional encoding as in \cite{liu2020learningencodepositiontransformer}.

Overall, the use of \ac{CNN}s on sinogram data is a convenience solution and the use of Transformers on projection data is shoe-horning, rather than adapted use of the intrinsic properties of sinograms. We argue that an architecture which generalises reasonably for sinogram must use a \emph{data-structure} adapted to the manifold nature of sinogram data during training and inference, rather than relying on an implicit parametrisation of the \ac{NN} when trained on a fixed geometry or on a unreasonably large model that could learn it from the data only. To address this, we propose to use the graph data-structure to represent the sinogram and its acquisition geometry, and create a novel architecture, the \ac{GLM}.  that combines traditional CNNs and \ac{GNN} to realise sinogram-to-sinogram mappings.

\section{Acquisition Geometries in CT}
 A typical setup in \ac{CT} involves measurements from multiple source positions, for which each of these have associated readouts from the detector elements. In a 2D setting, there is a line detector, this means that the sinogram will be a collection of 1D array for each source position. Likewise, in 3D, the sinogram consists of 2D arrays of detector elements for each source position. The aforementioned readouts can after simple preprocessing be interpreted as the line integrals of the object's attenuation along lines coming from the source.

The acquisition geometry is a precise description of the source/detector's motion relative to the scanned object. Mathematically, a sinogram can be seen a sampling of a real-valued function defined on some manifold of lines $\mathcal{M}$ that is described by the acquisition geometry. Being a manifold, $\mathcal{M}$ has a specific geometric structure that differs from the $\mathbb{R}^d$. Hence, signal processing of sinograms, like denoising or inpainting, should account for this geometric structure.

\begin{figure*}[t]
    \centering
    \begin{tikzpicture}
        \draw (0,0) ellipse (2cm and 0.6cm);
        \draw (0,4) ellipse (2cm and 0.6cm);
        
        \draw (-2,0) -- (-2,4);
        \draw (2,0) -- (2,4);
        
        \draw [decorate, decoration = {brace},line width=1pt] (-2.5,0) -- (-2.5,4);    
        \node at (-3,2) {$\mathcal{D}$};
         \draw[-stealth,line width=1pt] ($(-2,4.3)$) arc [start angle=180, end angle=0, x radius=2cm, y radius=0.6cm];
         \node at (2,5) {$S^1$};
    \end{tikzpicture}
    \begin{tikzpicture}
        \node at (0,0) {};
        \draw[->] (0,2.5) -- (2,2.5);
    \end{tikzpicture}
    \begin{tikzpicture}
        \node at (0,0) {};
        \draw[postaction={decorate},
          decoration={
              markings,
              mark = at position 0.07 with { \fill[black] circle (2pt); },
              mark = at position 0.19 with { \fill[black] circle (2pt); },
              mark = at position 0.30 with { \fill[black] circle (2pt); },
              mark = at position 0.45 with { \fill[black] circle (2pt); },
              mark = at position 0.62 with { \fill[black] circle (2pt); },
              mark = at position 0.74 with { \fill[black] circle (2pt); },
              mark = at position 0.89 with { \fill[black] circle (2pt); },
          }, dotted] (0.3,0) ellipse (2cm and 0.6cm);
        \draw[postaction={decorate},
          decoration={
              markings,
              mark = at position 0.07 with { \fill[black] circle (2pt); },
              mark = at position 0.19 with { \fill[black] circle (2pt); },
              mark = at position 0.30 with { \fill[black] circle (2pt); },
              mark = at position 0.45 with { \fill[black] circle (2pt); },
              mark = at position 0.62 with { \fill[black] circle (2pt); },
              mark = at position 0.74 with { \fill[black] circle (2pt); },
              mark = at position 0.89 with { \fill[black] circle (2pt); },
          }, dotted] (0.3,4) ellipse (2cm and 0.6cm);
        
        \draw[dashed] (-1.7,0) -- (-1.7,4);
        \draw[dashed] (2.3,0) -- (2.3,4);
    \end{tikzpicture}
\caption{Representing 2D tomography geometry: the source moves on the circle $S^1$ and the detector $\mathcal{D}$ is a line detector. The sampling on $S^1$  commonly varies in applications, whereas the sampling on $\mathcal{D}$ only varies when the detector is replaced. This is illustrated by the dotted curved with large black dots at given angular positions and fixed dashed lines on $\mathcal{D}$. A CNN-based approach assumes that training data has the same sampling of $S^1$ and $\mathcal{D}$ in training and inference, which is usually not the case, thus leading to poor generalisation.}
\label{label:2d_manifold}
\end{figure*}
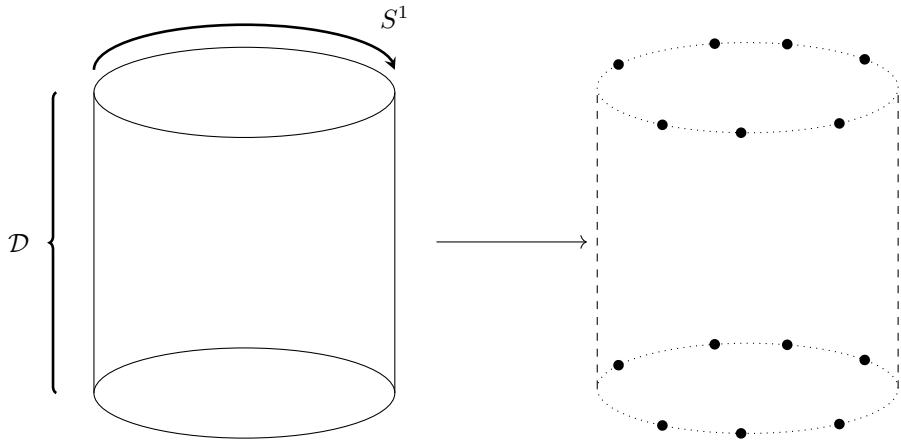

As explained in the introduction, the existing approaches for sinogram-to-sinogram mappings are not parametrised by the acquisition geometry, although some try to learn expected properties of the sinogram rather than explicitly using an encoding of its data manifold. This leads to poor generalisation even for small variations in the acquisition geometry. We expect such variations to occur in many tomographic applications as, although the detector is not changed in between sampling, there are always minor discrepancies between acquisition geometries. For instance, in medical CT, the angular resolution changes depending on the clinical question. To address this problem, we propose to structure the acquisition geometry using a graph data structure and then process the sinogram with a dedicated NN. We begin with a mathematical formalisation of the acquisition geometry to then move towards the definition of the GNN. 

For circular acquisition geometries, the source rotates on a circular path around the object. To describe this mathematically, let $S^1$ be the 1-dimensional sphere, and $\mathcal{D}$ represents the (non-discretised) detector. The manifold $\mathcal{M}$ of lines from the source to points on the detector is
\begin{equation}
\label{eq:manifold}
    \mathcal{M} \subset S^1 \times \mathcal{D}.
\end{equation}
The sinogram is now a function  
\begin{equation}
    F : \mathcal{M} \rightarrow \mathbb{R}.
\end{equation}
Hence, $\mathcal{M}$ is a product space, which means that the sampling on $S^1$ (source positions) and $\mathcal{D}$ (detector pixels) are commonly independent, as illustrated in Fig.~\ref{label:2d_manifold}. With this formulation in mind, we remark that using CNNs on sinogram data will lead to underperformance for two reasons. First, it does not make use of the geometry during inference, which seriously undermines generalisation. Second, the CNN learn to filter measurements and aggregate them with the same convolution kernels, which increases the number of required parameters without necessarily improving model capacity. 

Instead of processing a sinogram as if it were an image, we propose to filter it using a spatial convolution kernel in the detector dimension and a graph convolution kernel in the source position dimension. We aim to decorrelate the learning process for the two dimensions and build an architecture more resistant to perturbations in the acquisition geometry.

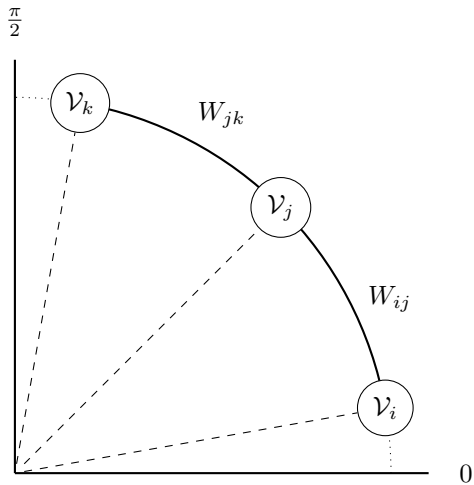
\begin{figure}[ht]
	\centering
	\begin{tikzpicture}
		\coordinate (centre) at (0,0);   
		\node (centre)  {};
		\def\R{5.0}                 
		\def\angStart{10}           
		\def\angEnd{80}            
		
		\draw[thick] (centre)++(\angStart:\R)
		arc[start angle=\angStart, end angle=\angEnd, radius=\R];
		
		\draw[dotted] (centre)++(0:\R)
		arc[start angle=0, end angle=\angStart, radius=\R];
		
		\draw[dotted] (centre)++(80:\R)
		arc[start angle=\angEnd, end angle=90, radius=\R];
		
		\node at (0, \R +1) {$\frac{\pi}{2}$};
		\node at (\R +1, 0) {$0$};
		
		\draw[thick] (0,0) -- (0, \R +0.5);
		\draw[thick] (0,0) -- (\R +0.5, 0);
		
		\foreach \k/\name in {0/$\mathcal{V}_i$,1/$\mathcal{V}_j$,2/$\mathcal{V}_k$}{
			\pgfmathsetmacro{\ang}{\angStart + \k*(\angEnd-\angStart)/2}
			\node[circle, draw=black, fill=white, inner sep=3pt,minimum size=10pt]
			(P\k) at ($(centre)+(\ang:\R)$) {\name};
			\draw[dashed] (centre) -- (P\k);
		}
		
		\node (W0) at ($(centre)+(25:\R+0.5)$) {$W_{ij}$};
		\node (W0) at ($(centre)+(60:\R+0.5)$) {$W_{jk}$};
		
	\end{tikzpicture}
	\caption{Source positions on the first quadrant of a polar grid. For each node, the feature vector corresponds to the data acquired at the associated source position.}
	\label{fig:acquisition_geometry}
\end{figure}

\section{Graphs, Convolutions, and Neural Networks }
Sinogram data lies on $\mathcal{M}$, a line manifold, which we propose to structure with a graph. 

\subsection{Relation between Acquisition Geometry and Graph Representation}
We define $G$, the graph associated to an acquisition geometry $\mathcal{M}$ in \eqref{eq:manifold}, as 
\begin{equation}
	G = (\mathcal{V}, \mathcal{E}, W)
\end{equation}
Where $\mathcal{V}$ is the set of nodes (the finite set of source positions obtained by sampling on $S^1$),  $\mathcal{E}$ the set of edges between the nodes and $W$ the weighted adjacency matrix. We note $n = |\mathcal{V}|$ the number of nodes. In our approach, we propose that the nodes encode measurements, whilst edges encode geometry as the nodes are connected only to their immediate neighbourhood, in a geometry-dependent way. As for the weights of $W$, there is no closed form expression of the relation between two measurements because it depends on $\mathcal{M}$ and the scanned object' properties. Although statistical tests could be used to determine the similarity between consecutive measurements a-priori or one could devise a learned weighting strategy, we leave this consideration out of the paper's scope deliberately and focus only on encoding the acquisition geometry in the edge's weighting scheme which we define in \eqref{eq:weighting_scheme}. 
\begin{align}
	\label{eq:weighting_scheme}      
	W&(A_i, A_j) = \cos(A_j - A_i) 
\end{align}
An approximation of \eqref{eq:weighting_scheme} for densely sampled measurements yields a one-valued constant weighting. We illustrate the relation between graph and geometry in Fig. \ref{fig:acquisition_geometry}. It is important to consider that this scheme cannot be applied as is to any acquisition geometries, even within the subset of circular geometries. For instance, for limited-angle CT, where a large number of consecutive measurements are missing, the connection of angles separated by a substantial angular difference is not straightforward, notwithstanding the fact that they are ``consecutive''. Alternatively, for parallel-beam CT, measurements separated by $\pi$ radians should be connected although they are far-away in the acquisition geometry, but the weight to apply to the edge in the graph is also not obvious. We underline that the encoding of the sinogram on a graph is a way to model the relation between measurements rather than a natural arrangement of the data, as could emerge in social networks or biology. It is a choice that stems from our understanding of line manifolds, but a choice nonetheless. 

\subsection{Convolution on Graphs, convolution on Grids}
\label{subsec:graph_convolution}
Equipped with a graph structure to represent the sinogram and the acquisition geometry, we will now explain how to use CNNs to filter the measurements and GNNs to aggregate them. CNNs are well studied and understood networks, but their graph counterparts are often less clear, especially in the field of imaging. As such, we fill focus on GNNs and explain the design of localized convolutional kernels on graphs based on \cite{defferrard_convolutional_2017} and \cite{smola_kernels_2003}.

The motivation for convolution on graphs stems from the need to have architectures that extract meaningful statistical patterns on graph data. As such, the term convolution corresponds to the ability to extract local information and combine it in a large-scale, hierarchical structure by composing layers. On grids, the convolution is a localised operation due to the finite size of the kernel and scale on the full data domain using translation. On graphs however, there is no unique definition of kernel translation as each node has a varying neighbourhood, so we can resort to the spectral understanding of the convolution. The convolution is a linear operator with complex exponential functions as eigenvectors. As such, the convolution diagonalizes (i.e it becomes a product) in the Fourier basis and the convolution of a signal by a kernel is equal to the inverse Fourier transform of the product of the Fourier transforms of the signal and the kernel. For graphs, the convolution is defined in the Fourier basis, using the eigenvectors of an operator that are also complex exponential functions, the graph Laplacian $L$, defined as  
\begin{equation}
	\label{eq:laplacian_definition}
	L = D - W.
\end{equation}
With $D$ the degree matrix and $W$ the adjacency matrix. For a signal $x:\mathcal{V} \to \mathbb{R}$, the action of $L$ is
\begin{equation}
	\label{eq:laplacian_action}
	(Lx)_i = \sum_{j \in \mathcal{N}_i} x_i - x_j.
\end{equation}
Where $\mathcal{N}_i$ is the neighbourhood of the graph node labelled $i$. Eq. \eqref{eq:laplacian_action} indicates that the Laplacian has a local impact, limited to each node's neighbourhood. Noting $U = [u_0,...,u_{n-1}]^T$ the set of orthonormal eigenvectors of $L$ associated with $\Lambda = \mathrm{diag}([\lambda_0,...,\lambda_{n-1}])$ the eigenvalues (or frequencies) of the graph, the decomposition of a signal $x$ defined on a graph in the Fourier basis is $\hat{x} = U^Tx $. In the spectral domain, the convolution is an element-wise multiplication (Hadamard product) between the frequency filter and the Laplace transform of the input. As such, the convolution can be defined as 
\begin{equation}
	\label{eq:spectral_convolution}
	g_\theta * x = U g_\theta(\Lambda)U^Tx.
\end{equation}
There is a crucial detail hiding in \eqref{eq:spectral_convolution}. As the eigenvectors of $L$ depend on node labelling, defining $g_\theta$ as a function of $\Lambda$ makes it depend on the frequencies of the graph only, and ensures independence on node permutations (relabelling). In other words, the kernel depends on the graph, not on its representation. 

Although there is a closed-form expression of the spectral convolutions on a graph, the convolution depends on an eigendecomposition of $L$ which can be prone to errors and approximations, or not even be feasible for large graphs. As detailed in \ref{subsubsec:spectral_analysis}, there is a closed form expression of the eigenvectors of the Laplacian of the graph we use to model a densely sampled, circular geometry. However, the graph convolution as implemented in modern deep-learning frameworks do not use the eigendecomposition. As defined in \eqref{eq:spectral_convolution}, $g_\theta(\Lambda)$ is a diagonal matrix which is problematic because it depends on $n$ and it is not localised in space. To address this problem, \cite{hammond_wavelets_2009} proposed to approximate $g_\theta(\Lambda)$ as a polynomial of the eigenvalues
\begin{equation}
	\label{eq:polynomial_approximation}
	g_\theta(\Lambda) \approx \sum_{i=0}^{K} \theta_k \Lambda^k.
\end{equation}
This approximation holds due to the Weierstrass approximation theorem which states that every continuous function defined on a closed interval can be uniformly approximated as closely as desired by a polynomial function. As the eigenvalues of the graph Laplacian lie in $[0,2]$ (see item v of Lemma 1.7 of chapter 1 of \cite{chung_spectral_1997}), a function of the eigenvalues can be approximated with a $K$-order polynomial, which \cite{hammond_wavelets_2009} proposed to be the Chebyshev polynomials. As these polynomials have eigenvalues in $[-1,1]$, the eigenvalues of the Laplacian are rescaled using $\tilde{\Lambda} = \Lambda - I$, with $I$ the identity matrix. Noting $\tilde{L} = L -I$ and integrating this approximation in \eqref{eq:spectral_convolution} yields
\begin{equation*}
	\label{eq:kth-approximation}
	g_\theta(\Lambda)  \approx U \sum_{i=0}^{K} \theta_k' \tilde{\Lambda}^k U^T x 
	 \approx \sum_{i=0}^{K} \theta_k' U \tilde{\Lambda}^k U^T x.
\end{equation*}
As $U \tilde{\Lambda}^k U^T = (U \tilde{\Lambda} U^T)^k = \tilde{L}^k$, a $K$-th order approximation of $g_\theta(\Lambda)$ is
\begin{equation}
	\label{eq:kernel_approximation}
	g_\theta(\Lambda) \approx \sum_{i=0}^{K} \theta_k' \hat{L}^k.
\end{equation}
Truncating \eqref{eq:kernel_approximation} to degree $K=1$ yields
\begin{equation}
	\label{eq:truncated_kernel_approximation}
	g_\theta(\Lambda) \approx \theta_0' + \theta_1'(L-I).
\end{equation}
Eq. \eqref{eq:truncated_kernel_approximation} highlights that graph convolution is a localised operation (cf \eqref{eq:laplacian_action}) and that stacking $k$ convolutional layers will apply $k$ times $L$ on the input signal which leads to instability when building deep neural networks. To address this \cite{kipf_semi-supervised_2017} proposed to use a normalization trick using the normalized graph Laplacian $L_{\text{norm}}$. Noting $D^{-\frac{1}{2}}$ the element-wise inverse square root of $D$ and $I$ the identity matrix, the definition of $L_{\text{norm}}$ is 
\begin{equation}
	\label{eq:lnorm}
	L_{\text{norm}} = I - D^{-\frac{1}{2}}WD^{-\frac{1}{2}}.
\end{equation}
which, combined to \eqref{eq:truncated_kernel_approximation} yields
\begin{equation}
	\label{eq:lnorm_reformulated}
	g_\theta \approx \theta_0' - \theta_1'D^{-\frac{1}{2}}WD^{-\frac{1}{2}}.
\end{equation}
\cite{kipf_semi-supervised_2017} uses parameter sharing to approximate $g_\theta*x$ as 
\begin{equation}
	\label{eq:graph_convolution_kipf}
	g_\theta*x \approx \theta(\tilde{D}^{-\frac{1}{2}}\tilde{W}\tilde{D}^{-\frac{1}{2}}) x.\\
\end{equation}
Where $\tilde{A} = A + I$ and $\tilde{D}_{ii} = \sum_j \tilde{A_{ij}}$. 
\eqref{eq:graph_convolution_kipf} is at the core of many implementations as it relies on the sparse to dense matrix multiplication $\tilde{A}x$ and on the diagonal matrix $\tilde{D}$. The local aggregation of node features by the Laplacian is referred to as \emph{message passing} and avoids the explicit computation of graph Fourier transform. 

\subsection{Graph Spectral Analysis}
\label{subsubsec:spectral_analysis}
We recall that the graph structure we use to model sinogram geometries acquired by dense sampling on circular paths is a graph that connects one measurement to its immediate neighbours with cosine weighting. In this case, the weighting scheme designed yields a constant, one-valued weight for each edge. Under these two assumptions, we can derive the that graph is cyclic, and its Laplacian is a circulant square $n \times n$ matrix with associated polynomial $f(x)$ so that
\begin{equation}
	f(x) = 2 - (x + x^{n-1}).
\end{equation}
Noting $w = \exp{(\frac{2i\pi}{n})}$, the Laplacian has eigenvectors
\begin{equation}
	u_j = \frac{1}{\sqrt{n}}(1, w^{j}, w^{2j}, ..., w^{j(n-1)})^T \forall j \in [0, n-1].
\end{equation}
and eigenvalues 
\begin{equation}
	\lambda_j = 2 - (w^{j} + w^{j(n-1)}).
\end{equation}
Although we have a closed form expression of the eigendecomposition of $L$, the message-passing neural networks sidestep their use altogether. In this paper, we will not try to compete with well-established sparse implementation of the spectral convolution, but we suppose that this knowledge could be used to regularise training by selecting which Fourier modes to amplify or dampen.

\section{Neural Network Design}
\label{subsec:nn_design}
We have defined the graph that we use to structure the sinogram data and its associated acquisition geometry. We also explained the graph convolution principle and how it depended on the message passing mechanism to propagate signal embeddings through the graph. We can now define the \ac{GLM} module, a NN building block realising a sinogram-to-sinogram mapping.

\subsection{\Ac{GLM} Modules}
\begin{figure}
	\centering
	\begin{tikzpicture}
		\node (Y0) at (0,0) {$Y$};
		\node (Y1) at (2,0) {$Y'$};
		\node (mp) at (3,0) {$\bigoplus$};
		\node (G) at  (3, -1) {$G$};
		\node (Y2) at (4,0) {$Y''$};
		\node (Y3) at (6,0) {$\hat{Y}$};
		\draw[->] (Y0) -- (Y1) node[midway,below] () {$f_{0}$}; 
		\draw[->] (Y1) -- (mp);
		\draw[->] (G) -- (mp);
		\draw[->] (mp) -- (Y2);
		\draw[->] (Y2) -- (Y3) node[midway,below] () {$f_{1}$};
	\end{tikzpicture}
	\caption{A \ac{GLM} module architecture. $f_{0}$ is a plain convolutional layer and $f_1$ is a residual convolutional layer, both followed by a ReLU activation function. The kernels dimension is equal to the detector's dimension. The $\bigoplus$ operator designates the $G$-dependent message-passing step.}
	\label{fig:glm_module}
\end{figure}
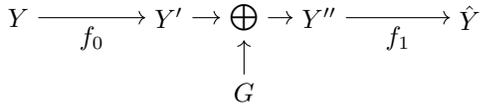

We introduce \ac{GLM}, a neural network module that combines the spatial convolution to filter measurements and the graph convolution to aggregate them. As explained in \ref{subsec:graph_convolution}, graph convolution is a local operation, in the sense that applying once the graph Laplacian on the input allows signal propagation in the immediate neighbourhood of each point. As such, to propagate information to distant nodes, we stack \ac{GLM} modules. Noting $f_0$ and $f_1$ two NN and $\bigoplus$ the message passing step, the mapping realised by the \ac{GLM} is defined as 
\begin{equation}
	\hat{Y}_i = f_1 (\bigoplus_{j \in \mathcal{N}_i} f_0(Y) )
\end{equation}
With $Y$ the sinogram and $i$ its i-th measurement. The choice of neural networks for $f_0$ and $f_1$ is up to the user. In this paper, we choose two CNN, $f_0$ being a plain convolutional one and $f_1$ a residual one. Noting $c$ the number of kernels of a layer, our \ac{GLM} module realises the channel mapping $c_{\text{in}} \to c_{\text{out}} \to c_{\text{out}}$. We illustrate a module in Fig. \ref{fig:glm_module}

A \ac{GLM} module aggregates information from its immediate neighbourhood. To extend it, one can simply stack modules. This highlights the flexibility of \ac{GLM}-based networks, as the final neighbourhood considered depends on the number of modules which is independent from a chosen kernel size, unlike for the CNN. As such, we decorrelate how we filter each measurements to how we aggregate them. Moreover, a \ac{GLM} module and a similar CNN module share the the same kernel size $S$. But, as \ac{GLM} kernels have one fewer dimension than the CNN's, they will have $S$ times fewer trainable parameters than an equivalent CNN. This does not mean that the memory cost is $S$-times fewer, as the graph convolution still needs to load the adjacency and degree matrices in memory. One key advantage of \ac{GLM} architectures however, is that all its trainable parameters are dedicated to signal processing rather than learning how to combine measurements. After passing through either network, the number of measurements and their dimension remain unchanged, as we do not downsample the measurements but rather use padding to preserve data dimensions.

\subsection{Computational Complexity}
\label{subsec:computational_complexity}
The computational complexity of a neural network is key to evaluate training time and costs. 
We note $P$ the number of detector pixels and $S$ the kernel size. For the \ac{GLM} module, there are two 1D-CNN evaluation and one neighbourhood aggregation. The complexity of a module is then  
\begin{equation}
	\label{eq:glm_complexity}
	n  P  S  c_{\text{in}}  c_{\text{out}} + 3n + n   P   S  c_{\text{out}}  c_{\text{out}} 
\end{equation}
The factor 3 comes from the fact that each node has two neighbours and a self loop. For an equivalent plain CNN module, there are two 2D-CNN evaluations, which has a complexity of he complexity of
\begin{equation}
	\label{eq:cnn_complexity}
	n  P  S^2  c_{\text{in}}  c_{\text{out}} + n  P S^2  c_{\text{out}}  c_{\text{out}} 
\end{equation}
Simplifying \eqref{eq:glm_complexity} and \eqref{eq:cnn_complexity} shows that the neighbourhood aggregation adds a fixed cost to the evaluation whereas the complexity scales proportional to the kernel dimension for the CNN. This indicates that the graph-based approach scales much better than the CNN baseline.

\subsection{Implementation Details}
We want to realise a sinogram-to-image mapping. To do so, we compose a sinogram-to-sinogram mapping with a pseudo inverse of $\mathcal{A}$ and with an image-to-image mapping. In this section, we explain how we will define two sinogram-to-sinogram mappings that rely on different aggregation mechanisms (CNN and \ac{GLM}) whilst still being comparable.

 It is crucial to note that the aggregation of measurements depends on the number of modules stacked for \ac{GLM}, whereas it depends on the kernel size for the CNNs. As such, we decide to use 3 modules with a kernel of size 7. In this setting, the CNNs should be able to aggregate information 3 rows away and the \acp{GLM} should gather information three hops away. This highlights the flexibility of the \ac{GLM}, as its neighbourhood depends on the number of modules which is independent from its kernel size, unlike for the CNN.   
 
 To avoid confusion, we will refer by \ac{GLM} and $\mathrm{CNN}$ to the networks obtained by stacking $3$ \ac{GLM} modules and $3$ CNN modules without graph convolution, respectively. It is important to note that both modules have the same number of layers and are composed of a plain CNN and as ResNet. In terms of channel mapping, \ac{GLM} and $\mathrm{CNN}$ do the mapping $1 \to c \to c \to 1$. In order to distinguish between networks having the same architecture but a different number of channels, we refer to the networks as $\mathrm{GLM}{\text -}c$ and $\mathrm{GLM}{\text -}c$ from now on. 

As we are interested in sinogram-to-sinogram mapping networks, we fix the two other operators. As pseudo inverse, we use the Filtered Back-projection with a Ram-Lak filter. For the image-to-image mapping, we use a plain convolutional network $\Gamma$ which is identical for the two pipelines. A \ac{GLM} module is implemented by modifying the GCNConv module of \cite{kipf_semi-supervised_2017} using the PyGeom \cite{fey_fast_2019} package. The spatial convolutions are implemented using PyTorch \cite{paszke_pytorch_2019}. We first pretrained CNN and \ac{GLM} on a sinogram-to-sinogram autoencoding task for 1 epoch. Then, we trained the reconstruction pipelines using Adam optimiser for 40 epochs with a learning rate of $5.10^{-5}$. All training was done with a batch size of 8.

\section{Experiments}
We want to assess how a sinogram-to-sinogram mapping network made by stacking \ac{GLM} layers and aware of the acquisition geometry performs against a plain CNN, traditionally used in learned reconstruction frameworks. We train \ac{GLM} and $\mathrm{CNN}$ for 16 and 24 channels against a fixed, densely sampled acquisition geometry on the training dataset. We then evaluate them on the test dataset, first with the same geometry and then on an angularly subsampled acquisition geometries. We finally conduct a study on the scalability of either approaches. Our quantitative results are reported using the \ac{SSIM} and \ac{PSNR} metrics. For the \ac{SSIM}, we normalise the images and convert them to 8-bits integer, and use the \texttt{pytorch-ssim} python package for the computation.

\begin{figure*}[t]
	\centering
	\subfloat[]{\includegraphics[width=2.0in]{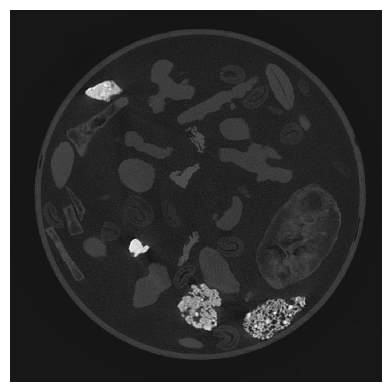}%
		\label{fig:conv2d_fixed_geom}}
	\hfil
	\subfloat[]{\includegraphics[width=2.0in]{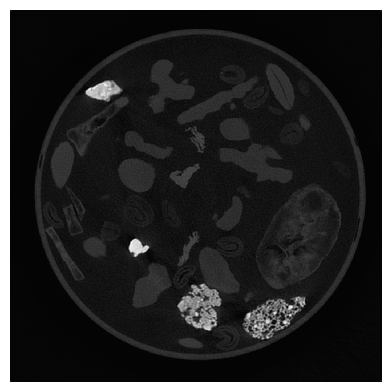}%
		\label{fig:GLM_fixed_geom}}
	\hfil
	\subfloat[]{\includegraphics[width=2.0in]{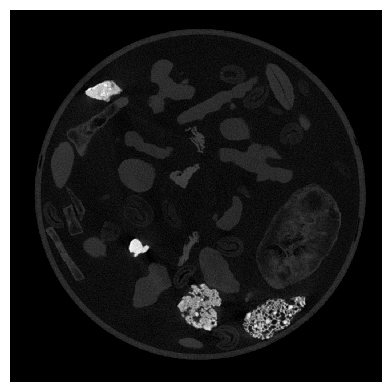}%
		\label{fig:target_fixed_geom}}
	\caption{Comparison of the same slice reconstructed using the baseline CNN approach \ref{fig:conv2d_fixed_geom}, the proposed graph-based approach \ref{fig:GLM_fixed_geom} and the target \ref{fig:target_fixed_geom}. We can see that the CNN-based approach produces a grey background compared to the target and the \ac{GLM}.}
	\label{fig:qualitative_results_fixed_geometry}
\end{figure*}

\subsection{Dataset}
We use the 2DeteCT \cite{kiss_2detect_2023} dataset which contains 5000 2D slices sampled from 110 objects with high natural variability in shape and density. In the original dataset, each sample consists in three different quadruplets: the flat and dark field sinograms, the sinogram and the high-quality reconstruction obtained by Nesterov accelerated gradient descent. There are three of these quadruplets per-sample, each corresponding to a different acquisition scenario. ``Mode1'' sinograms have a low photon count (tube power of 3W), ``Mode2'' a normal photon count (tube power of 90W) and ``Mode3'' are laden with inflicted beam-hardening artefacts. In this paper, we are only interested in Mode2 sinograms. For computational efficiency, we pre-process each sinogram using the LION \cite{biguri_lion_nodate} toolbox. We refer hereafter to a sample being a pair of a pre-processed Mode2 sinogram ($3600\times956$ pixels array) and its associated $1024\times1024$ pixels reconstruction. To split the dataset between training and testing, we divide the $5.000$ original slices per-object rather than divide the slices directly to avoid data-leakage. As such, there are $3.930$, $600$ and $470$ slices in the training, validation and testing datasets.

\subsection{Performance against the Training Acquisition Geometry}
We train the four netwoks on the fully sampled acquisition geometry ($3600\times956$ pixels sinograms) and evaluate on the test dataset with the same geometry. The networks are trained to reconstruct a high-quality image from a Mode2 sinogram by Mean-Square Error (MSE) minimisation. We report the \ac{SSIM} and \ac{PSNR} of the reconstructed image against the target reconstruction, and the number of trainable parameters of each network in Table \ref{tab:angular_subsampling1}. 

\begin{figure}
	\centering
	\begin{tikzpicture}
		\begin{axis}
			
			\addplot[solid] coordinates {
				(0, 1.00)(1, 0.00)(2, 0.01)(3, 0.01)(4, 0.01)(5, 0.02)(6, 0.02)(7, 0.03)(8, 0.04)(9, 0.05)(10, 0.06)(11, 0.07)(12, 0.07)(13, 0.08)(14, 0.08)(15, 0.08)(16, 0.07)(17, 0.07)(18, 0.06)(19, 0.05)(20, 0.05)(21, 0.04)(22, 0.03)(23, 0.03)(24, 0.02)(25, 0.02)(26, 0.02)(27, 0.01)(28, 0.01)(29, 0.01)(30, 0.01)(31, 0.01)(32, 0.01)(33, 0.01)(34, 0.01)(35, 0.01)(36, 0.01)(37, 0.02)(38, 0.02)(39, 0.02)(40, 0.02)(41, 0.02)(42, 0.02)(43, 0.02)(44, 0.02)(45, 0.02)(46, 0.02)(47, 0.02)(48, 0.02)(49, 0.01)(50, 0.01)(51, 0.01)(52, 0.01)(53, 0.01)(54, 0.01)(55, 0.01)(56, 0.00)(57, 0.00)(58, 0.00)(59, 0.00)(60, 0.00)(61, 0.00)(62, 0.00)(63, 0.00)(64, 0.00)(65, 0.00)(66, 0.00)(67, 0.00)(68, 0.00)(69, 0.00)(70, 0.00)(71, 0.00)(72, 0.00)(73, 0.00)(74, 0.00)(75, 0.00)(76, 0.00)(77, 0.00)(78, 0.00)(79, 0.00)(80, 0.00)
			};
			\addplot[dashed] coordinates {
				(0, 0.00)(1, 0.00)(2, 0.00)(3, 0.08)(4, 0.46)(5, 0.33)(6, 0.06)(7, 0.04)(8, 0.02)(9, 0.01)(10, 0.01)(11, 0.02)(12, 0.02)(13, 0.03)(14, 0.05)(15, 0.06)(16, 0.08)(17, 0.10)(18, 0.10)(19, 0.10)(20, 0.10)(21, 0.08)(22, 0.07)(23, 0.06)(24, 0.05)(25, 0.04)(26, 0.03)(27, 0.03)(28, 0.02)(29, 0.02)(30, 0.01)(31, 0.01)(32, 0.01)(33, 0.01)(34, 0.01)(35, 0.01)(36, 0.01)(37, 0.01)(38, 0.01)(39, 0.01)(40, 0.01)(41, 0.02)(42, 0.02)(43, 0.02)(44, 0.03)(45, 0.03)(46, 0.03)(47, 0.03)(48, 0.03)(49, 0.02)(50, 0.02)(51, 0.02)(52, 0.01)(53, 0.01)(54, 0.01)(55, 0.01)(56, 0.01)(57, 0.00)(58, 0.00)(59, 0.00)(60, 0.00)(61, 0.00)(62, 0.00)(63, 0.00)(64, 0.00)(65, 0.00)(66, 0.00)(67, 0.00)(68, 0.00)(69, 0.00)(70, 0.00)(71, 0.00)(72, 0.00)(73, 0.00)(74, 0.00)(75, 0.00)(76, 0.00)(77, 0.00)(78, 0.00)(79, 0.00)(80, 0.00)
			};
			\addplot[dotted] coordinates {
				(0, 0.00)(1, 0.00)(2, 0.00)(3, 0.00)(4, 0.00)(5, 0.00)(6, 0.00)(7, 0.00)(8, 0.00)(9, 0.00)(10, 0.00)(11, 0.00)(12, 0.00)(13, 0.00)(14, 0.00)(15, 0.00)(16, 0.00)(17, 0.00)(18, 0.00)(19, 0.00)(20, 0.00)(21, 0.05)(22, 0.49)(23, 0.40)(24, 0.04)(25, 0.02)(26, 0.01)(27, 0.01)(28, 0.02)(29, 0.03)(30, 0.04)(31, 0.06)(32, 0.08)(33, 0.09)(34, 0.11)(35, 0.11)(36, 0.10)(37, 0.09)(38, 0.08)(39, 0.06)(40, 0.05)(41, 0.04)(42, 0.03)(43, 0.03)(44, 0.02)(45, 0.02)(46, 0.01)(47, 0.01)(48, 0.01)(49, 0.01)(50, 0.01)(51, 0.01)(52, 0.01)(53, 0.01)(54, 0.01)(55, 0.02)(56, 0.02)(57, 0.02)(58, 0.03)(59, 0.03)(60, 0.03)(61, 0.03)(62, 0.03)(63, 0.03)(64, 0.02)(65, 0.02)(66, 0.02)(67, 0.01)(68, 0.01)(69, 0.01)(70, 0.01)(71, 0.00)(72, 0.00)(73, 0.00)(74, 0.00)(75, 0.00)(76, 0.00)(77, 0.00)(78, 0.00)(79, 0.00)(80, 0.00)
			};
			\legend{Target,$\mathrm{GLM}{\text -}{24}$, $\mathrm{CNN}{\text -}{24}$}
			\addplot[mark = o, only marks] coordinates {(0, 1)(14, 0.08)(42, 0.02)};
			\addplot[mark = x, only marks] coordinates {(4, 0.46)(18, 0.1)(41, 0.02)};
			\addplot[mark = +, only marks] coordinates {(22, 0.49)(34, 0.11)(61, 0.03)};
		\end{axis}
	\end{tikzpicture}
	\caption{Grayscale values histogram comparing of one sample of the test dataset. Just like for calculating the \ac{SSIM}, the images are first converted to 8-bits integers. For display purposes, we \emph{rescaled} the pixel counts in the $[0,1]$ interval and \emph{limited} the histogram to pixel values below $80$, has there were no pixel values above this threshold. We added marks at the locations of the peaks.}
	\label{fig:histogram}
\end{figure}
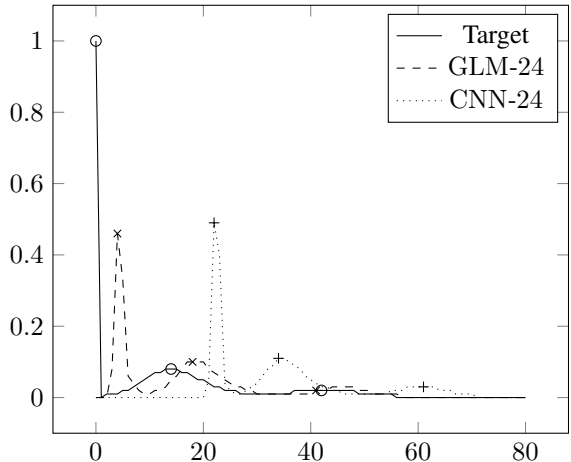

\begin{table}[h]
	\centering
	\begin{tabular}{cccr}
		Network & PSNR & SSIM & Parameters \\
		\toprule
		$\mathrm{CNN}{\text -}{16}$ & $40.46 \pm 1.93$ & $0.46 \pm 0.02$ & 39\,315\\ 
		$\mathrm{GLM}{\text -}{16}$ & $41.92 \pm 1.93$ & $0.47 \pm 0.01$ & 5\,673 \\ 
		$\mathrm{CNN}{\text -}{24}$ & $41.11 \pm 1.94$ & $0.46 \pm 0.02$ & 87\,171\\
		$\mathrm{GLM}{\text -}{24}$ & $40.39 \pm 1.93$ & $0.62 \pm 0.05$ & 12\,537\\
	\end{tabular}
	\caption{Quantitative metrics for the two learned reconstruction pipelines on the test dataset with the highly-sampled geometry. Note that the $7$ times increase in parameters between CNN and \ac{GLM} relates to the kernel size.}
	\label{tab:angular_subsampling1}
\end{table}
Beginning with an intra-module comparison, we see that nearly doubling the number of parameters of the CNN leads to a 0.6 dB PSNR increase and no SSIM improvement. For the \ac{GLM} however, this increase in model parameters yields a substantial 0.16 SSIM percentage point (pp) increase. This hints that \ac{GLM}'s trainable parameters are focused on improving image quality. Then, following up with an extra-module comparison, we can see that a CNN with fifteen times the number of parameters of a \ac{GLM} yields a worst performance in both metrics. This is a strong evidence that neural networks aware of the underlying data structure are better candidates than large models, challenging the ``the more, the better'' rule often followed in deep-learning. In practice, the \ac{SSIM} improvement observed translates into images of which histograms match the target reconstruction's better. Consider Fig. \ref{fig:histogram}. For the target reconstruction, most pixel values are concentrated at $0$, with two peaks around $15$ and $42$. For the images inferred by the learned reconstruction networks, we observe the same number of peaks, but with their position shifted and their height changed. For the \ac{GLM}, most values are concentrated around $4$, with two other peaks at $18$ and $41$. For the CNN, most values are concentrated around $22$, with two other peaks at $34$ and $61$. This qualitative measurement hints that \ac{GLM} produces images closer to the original ones as its first peak in pixel density is closer to the target's one, but also with better contrast, as the distance between first and second peak is larger than the CNN's. Finally, the CNN has the bulk of its pixel values around 22, which indicates a greyer background. This observation is backed by the Fig. \ref{fig:qualitative_results_fixed_geometry}, where a grey background can be seen inside the container of Fig. \ref{fig:conv2d_fixed_geom}.

\begin{figure*}[t]
	\centering
	\begin{tikzpicture}[scale=1]
		\begin{axis}[
			name=PSNR,
			xlabel=Angular Subsampling Factor,
			ylabel=PSNR,
            xtick={1,2,3,4,5,6,7,8,9,10},
			]
			\addplot[    
			dashed, mark = x
			] coordinates {
				(1, 40.46)(2, 37.54)(3, 35.62)(4, 34.34)(5, 33.42)(6, 32.70)(7, 32.08)(8, 31.54)(9, 31.06)(10, 30.64)
			};
			\addplot[    
			solid, , mark = x
			] coordinates {
				(1, 41.92)(2, 39.03)(3, 37.52)(4, 36.47)(5, 35.67)(6, 34.97)(7, 34.32)(8, 33.75)(9, 33.26)(10, 32.83)
			};
			\addplot[    
			dashed, mark = +
			] coordinates {
				(1, 41.11)(2, 38.39)(3, 36.82)(4, 35.72)(5, 34.87)(6, 34.17)(7, 33.54)(8, 32.99)(9, 32.51)(10, 32.09)
			};
			\addplot[    
			solid, mark = +
			] coordinates {
				(1, 40.39)(2, 38.76)(3, 37.56)(4, 36.62)(5, 35.84)(6, 35.20)(7, 34.65)(8, 34.15)(9, 33.68)(10, 33.24)
			};
			
			\legend{$\mathrm{CNN}{\text -}{16}$, $\mathrm{GLM}{\text -}{16}$, $\mathrm{CNN}{\text -}{24}$, $\mathrm{GLM}{\text -}{24}$}
		\end{axis}
		
		\begin{axis}[
			name = SSIM,
			at={(PSNR.right of south east)}, anchor=left of south west,
			ylabel=SSIM,
            xtick={1,2,3,4,5,6,7,8,9,10},
			]
			\addplot[    
			dashed, mark = x
			] coordinates {
				(1, 0.47)(2, 0.42)(3, 0.39)(4, 0.37)(5, 0.35)(6, 0.34)(7, 0.32)(8, 0.31)(9, 0.30)(10, 0.29)
			};
			\addplot[    
			solid, , mark = x
			] coordinates {
				(1, 0.46)(2, 0.42)(3, 0.39)(4, 0.38)(5, 0.36)(6, 0.35)(7, 0.34)(8, 0.32)(9, 0.31)(10, 0.30)
			};
			\addplot[   
			dashed , mark = +
			] coordinates {
				(1, 0.46)(2, 0.42)(3, 0.39)(4, 0.38)(5, 0.36)(6, 0.35)(7, 0.34)(8, 0.33)(9, 0.31)(10, 0.31)
			};
			\addplot[   
			solid, mark = +
			] coordinates {
				(1, 0.62)(2, 0.59)(3, 0.57)(4, 0.55)(5, 0.53)(6, 0.52)(7, 0.50)(8, 0.49)(9, 0.47)(10, 0.46)
			};
			
		\end{axis}
		
	\end{tikzpicture}
	\caption{Evolution of the PSNR and SSIM against the angular subsampling. The two graphs share the same $x$ axis and the same legend. The solid lines represent the performance of the \acp{GLM} and the dashed lines the CNNs. The networks with $16$ kernels use a $\times$ marker and the ones with $24$ use a $+$ marker.}
	\label{fig:quantitative_performance}
\end{figure*}
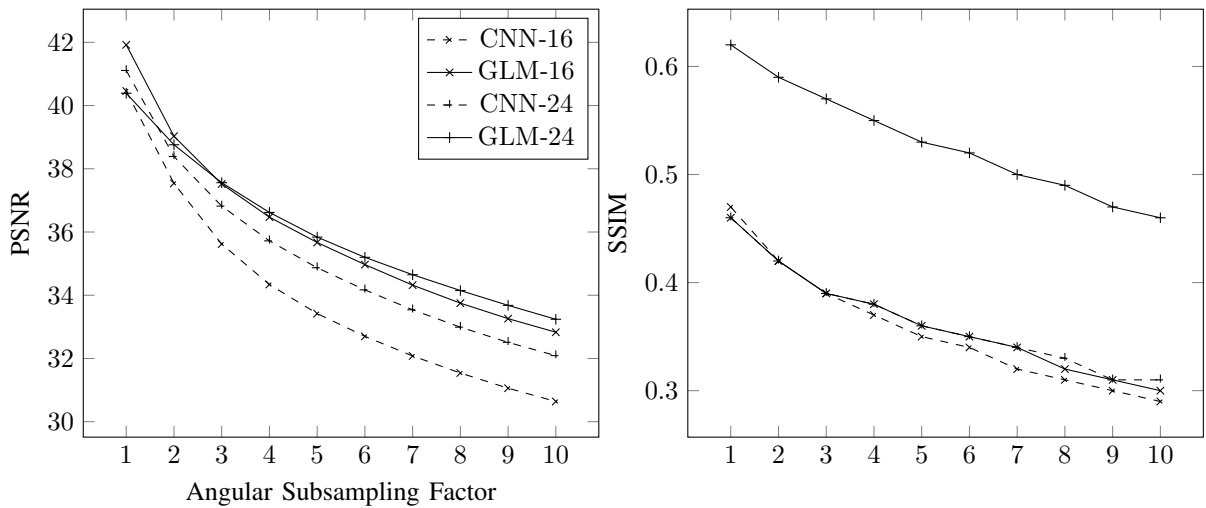

\subsection{Generalisation against Downsampled Acquisition Geometries}
After evaluating the networks against the geometry they were trained on, we want to understand how they generalise to unseen geometries and how they scale in terms of number of trainable parameters. To do so, we evaluated all four networks on $10$ angular subsampling scenarios and report the results in Fig. \ref{fig:quantitative_performance} and Table. \ref{tab:angular_subsampling2}.

Beginning with an intra-module comparison, we can see that increasing the number of parameters of a CNN improves the generalisation capability with regard to the PSNR. Indeed, $\mathrm{CNN}{\text -}{24}$ has a higher PSNR than $\mathrm{CNN}{\text -}{16}$ throughout the angular downsampling range and achieves a lesser drop. For the SSIM, the improvement is however marginal. For the \ac{GLM}, one can clearly see the strong impact of the number of parameters' increase for the SSIM in terms of absolute performance. When it comes to generalisation, we can observe that although $\mathrm{GLM}{\text -}{24}$ underperforms compared to $\mathrm{GLM}{\text -}{16}$, it actually generalises better than the latter from the angular subsampling of 4. This result is backed by the fact that the PSNR performance drops by $9.1$dB for $\mathrm{GLM}{\text -}{16}$ and only $7.2$dB for $\mathrm{GLM}{\text -}{24}$. These results when put together in an extra-module comparison show that using $\mathrm{GLM}{\text -}{16}$, a network with 15 times fewer parameters than $\mathrm{CNN}{\text -}{24}$, achieves better PSNR, similar SSIM and generalisation. Finally, as the generalisation performance and SSIM quality increase drastically when increasing the number of \ac{GLM} parameters, one can say that our architecture allocates resources better than the standard CNN.

\begin{table}[h]
	\centering
	\begin{tabular}{cccrrr}
		Network     & SSIM drop & PSNR drop\\
		\toprule
		$\mathrm{CNN}{\text -}{16}$ & $0.18$ pp & $9.8$ dB \\ 
		$\mathrm{GLM}{\text -}{16}$ & $0.16$ pp & $9.1$ dB \\ 
		$\mathrm{CNN}{\text -}{24}$ & $0.17$ pp & $9.0$ dB \\
		$\mathrm{GLM}{\text -}{24}$ & $0.16$ pp & $7.2$ dB \\
	\end{tabular}
	\caption{Quantitative metrics for the four learned reconstruction pipelines on the test dataset with downsampled acquisition geometries.}
	\label{tab:angular_subsampling2}
\end{table}

\subsection{Memory and Training Time Scalability}
\begin{figure*}[h]
	\centering
	\begin{tikzpicture}
		\begin{axis}[
			name=parameters,
			ylabel=Trainable Parameters ($\times 10^3$),
			xlabel=Number of Kernels,
			xtick={4,8,16,24,32,64},
			xticklabels={4,8,16,24,32,64},
			xmode=log,
			log basis x= 2,
			legend style={at={(0.05,0.85)},anchor=west}
			]
			\addplot[dashed, mark=+] coordinates {
				(4, 2.811)(8, 10.275)(16, 39.315)(24, 87.171)(32, 153.843)(64, 608.691)
			};
			\addplot[ solid, mark=+] coordinates {
				(4, 0.417)(8, 1.497)(16, 5.673)(24, 12.537)(32, 22.089)(64, 87.177)
			};
			
			\legend{CNN, GLM}
		\end{axis}
		
		\begin{axis}[
			name = Memory,
			at={(parameters.right of south east)}, anchor=left of south west,
			ylabel=Memory (Gb/sample),
			xtick={4,8,16,24,32,64},
			xticklabels={4,8,16,24,32,64},
			xmode=log,
			log basis x= 2,
            legend style={at={(0.05,0.85)},anchor=west}
			]
			\addplot[dashed, mark=+] coordinates {
				(4, 1.617782784)(8, 1.9465728)(16, 4.596504576)(24, 6.912141312)(32, 9.233268224)(64, 15.202925568)
			};
			\addplot[solid, mark=+] coordinates {
				(4, 2.26043904)(8, 2.993434112)(16, 4.0955264)(24, 5.0179456)(32, 6.888246272)(64, 12.392473088)
			};
            \legend{CNN, GLM}
		\end{axis}
		
	\end{tikzpicture}
	\caption{Evolution of the number of trainable parameters and per-sample memory against the number of kernels.}
	\label{fig:mem_vs_kernels}
\end{figure*}
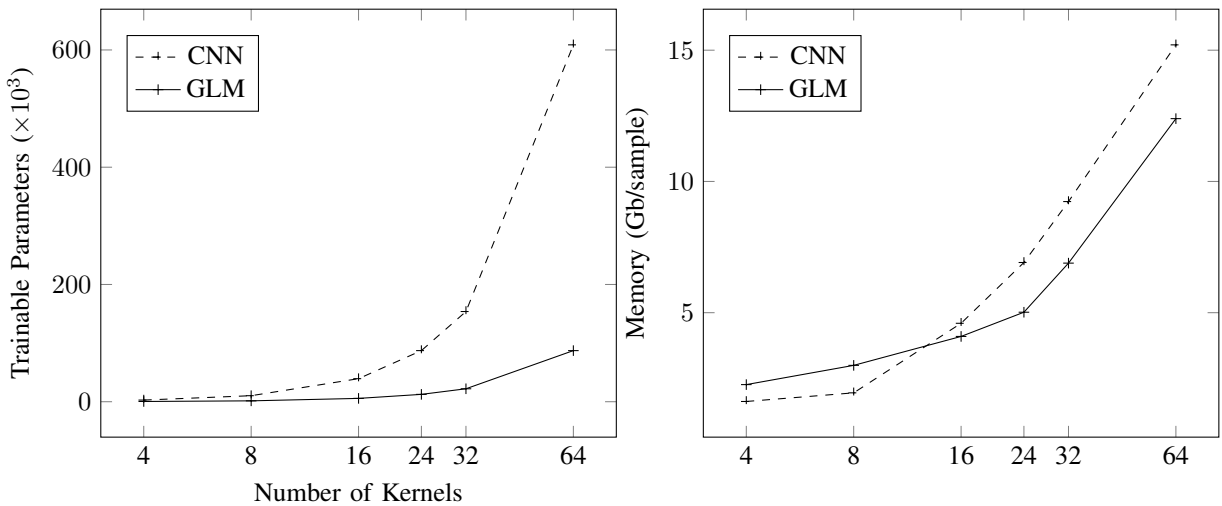

As discussed in section~\ref{subsec:computational_complexity}, we expect the \ac{GLM} modules to have a better scalability than the plain \ac{CNN} when it comes to training time and memory requirements. We begin by studying the evolution of the number of training parameters and per-sample memory requirements against the number of kernels. We present our results in Fig. \ref{fig:mem_vs_kernels}.

\begin{figure}[h]
	\centering
	\begin{tikzpicture}
		\begin{axis}[
			name=training,
			xlabel=Batch Size,
			ylabel=Batch Training Time (s),
            legend style={at={(0.05,0.85)},anchor=west}
			]
			\addplot[dashed, mark=+] coordinates {
				(2, 1.51)(4, 2.7)(6, 4.01)(8, 5.93)(10, 7.23)
			};
			\addplot[ solid, mark=+] coordinates {
				(2, 0.66)(4, 0.78)(6, 1.43)(8, 2.42)(10, 2.76)
			};
			
			\legend{CNN, GLM}
		\end{axis}
		
	\end{tikzpicture}
	\caption{Training time against Batch Size. \Acp{GLM} training time is not only smaller than the \acp{CNN}, it also scales better with regard to the batch size.}
	\label{fig:time_vs_batch}
\end{figure}
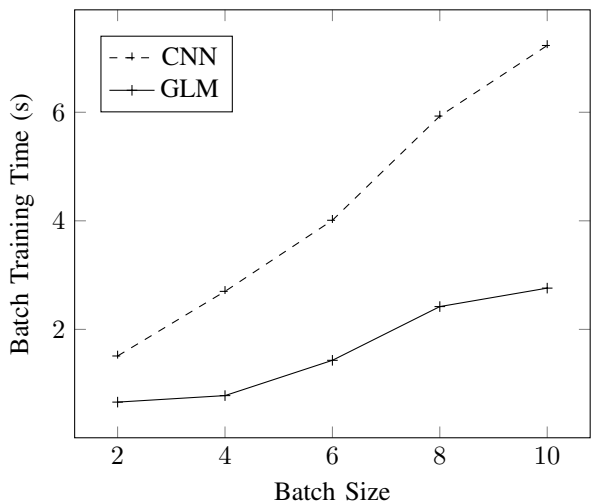

Crossing the networks' performance and the evolution of the number of trainable parameters shows an inflation for the \ac{CNN}, which scales poorly for no measurable improvement compared to the \ac{GLM}. For the per-sample memory, the story is different. As indicated in section~\ref{subsec:computational_complexity}, the \ac{GLM} come with fixed cost depending on the number of graph edges. This cost is offset as soon as there are more than 16 kernels, after which the \ac{CNN}'s memory requirements grows faster than the \ac{GLM}'s. As such, for the number of kernels used in our performance study, \ac{GLM} is less expensive to train and requires fewer parameters. These encouraging results are backed by the evolution of the training time against the batch size reported in Fig. \ref{fig:time_vs_batch}. To put it together, \ac{GLM} requires fewer parameters, memory and training time than its \ac{CNN} counterpart.

\section{Conclusion}
The majority of learned reconstruction frameworks rely on grid convolutions to process sinogram data. This approach, although convenient due to the off-the-shelf availability and  early arrival of \acp{CNN} in the field, has two major drawbacks, which are generalisation to unseen geometries and scalability. We identified that these two issues come from the fact that the \ac{CNN} are designed to process data on grids whereas sinogram data lie on line manifolds. To account for this specificity, recent research has looked towards Transformers, with mitigated success. The contribution of our approach is to introduce the use of the graph data structure to represent the sinogram and the acquisition geometry, and design a new neural network module, \ac{GLM}, that uses spatial convolution to extract sinogram features and graph convolution to aggregate them. Our experiences have shown that \ac{GLM} generalises better, requires fewer parameters and exhibits higher performance than its CNN counterpart, and claim that it is currently the best approach to tackle sinogram-to-sinogram mappings. We also propose research directions to enhance the approach even further.

We recall that we model the sinogram and acquisition geometry as a weighted, undirected graph. The graph's node features are the measurements read on the detector at a given source position, and the graph's weights depend on the distance between source positions. Although this approach has shown improvements compared to the baseline CNN, there are several questions left open. First, the graph representation, especially its connectivity and weighting, depends on the acquisition geometry. In this paper, we only focused on the circular, uniformly sampled cone-beam geometries. For these kind of ``well-behaved'' geometries, we believe that the Laplacian properties could be used to regularise training and maybe improve the computations, as underlined in \ref{subsubsec:spectral_analysis}, and that more edges could be added to connect measurements $\pi$ radians away. For other geometries, especially the helical ones, the graph weighting is yet to be defined to account for translational motion. As for the \ac{GLM} module itself, the message-passing mechanism produces a weighted sum of neighbouring acquisitions and must be permutation invariant. We hypothesise that using a GraphSAGE-based \cite{hamilton_inductive_2018} architecture could benefit the performance as it concatenates node features from the previous layer rather than aggregating them. Also, one could study learned graph weighting or graph-transformers \cite{bresson_residual_2018, rampasek_recipe_2023, rong_self-supervised_2020}.

\section*{Acknowledgments}
OÖ and EV acknowledge support from Swedish Energy Agency grant P2022-00286 and FORMAS grant 2022-00469.

\section{References Section}
\bibliographystyle{IEEEtran}
\bibliography{references.bib}

\vfill

\end{document}